\documentclass{article}
\usepackage[utf8]{inputenc}
\usepackage{graphicx}
\usepackage[version=3]{mhchem}
\newcommand{\nc}{\newcommand}
\nc{\nn}{\nonumber}

\begin{document}
\title{Matters Arising in ``Plasmon-driven carbon–fluorine (C(sp$^3$)–F) bond activation with mechanistic insights into hot-carrier-mediated pathways''}

\author{Yonatan Dubi$^{1,2}$, Ieng Wai Un$^3$, Joshua H.~Baraban$^1$, and Yonatan Sivan$^{2,3}$\\
Ben-Gurion University of the Negev, Israel}


\maketitle

In a recent paper~\cite{Halas-Nature-Catalysis-2020}, Robatjazi {\em et al.} demonstrate hydrodefluorination on Al nanocrystals decorated by Pd islands under illumination and under external heating. 
They conclude that photocatalysis accomplishes the desired transformation \ce{CH3F + D2 -> CH3D + DF} efficiently and selectively due to ``hot'' electrons, as evidenced by an illumination-induced reduction of the activation energy.

Although some of the problems identified in prior work by the same group~\cite{anti-Halas-Science-paper,R2R} have been addressed~\footnote{In particular, unlike the procedure employed in previous work by this group~\cite{Halas_Science_2018}, the reaction data was not normalized to different effective volumes in the current work, but rather by the {\em same} total catalyst mass. In that sense, the authors implicitly acknowledge the error we identified~\cite{anti-Halas-Science-paper,Y2-eppur-si-riscalda,R2R} in their previous work~\cite{Halas_Science_2018}.}, scrutiny of the data in~\cite{Halas-Nature-Catalysis-2020} raises doubts about both the methodology and the central conclusions. First, we show that the thermal control experiments in~\cite{Halas-Nature-Catalysis-2020} do not separate thermal from "hot electron" contributions, and therefore any conclusions drawn from these experiments are invalid. We then show that an improved but still non-ideal thermal control implies that the activation energy of the reaction does not change, and that an independent purely thermal calculation (based solely on the sample parameters provided in the original manuscript~\cite{Halas-Nature-Catalysis-2020}) explains the measured data perfectly. For the sake of completeness, we also address technical problems in the calibration of the thermal camera, an unjustifiable disqualification of some of the measured data, as well as concerning aspects of the rest of the main results, including the mass spectrometry approach used to investigate the selectivity of the reaction, and claims about the stoichiometry and reaction order. All this shows that the burden of proof for involvement of hot electrons has not been met.

\section{extraction of ``hot'' electron contribution 
(Fig.~6a)}
\subsection{Treatment of the thermocatalysis control data}

One of the central results of~\cite{Halas-Nature-Catalysis-2020} is shown in its Fig.~6a; it purportedly demonstrates that the activation energy decreases under illumination by plotting the ``hot'' carrier contribution to the rate vs.~the (intensity-dependent) surface temperature and performing Arrhenius fits. 

In order to explain the possible flaws in Fig.~6a, we must first explain the procedure used to construct it, as described in the manuscript. First, the reaction rate is measured in the dark ($R_{dark}$). The sample is then illuminated and the reaction rate, $R_{illum}$, is measured as a function of $T_S$ again.

However, $R_{illum}$ is \emph{not} what is plotted in Fig.~6a. Instead, to isolate the photocatalysis (i.e., the ``hot'' electron) contribution to the reaction rate under illumination, denoted by $R_{photo}$
, the authors take 
the difference between the reaction rate under 
illumination and the reaction rate in the dark, i.e., $R_{photo}(T_S) = R_{illum}(T_{S}) - R_{dark}(T_{dark})$ (see~\cite[SI, p. 30]{Halas-Nature-Catalysis-2020}). The temperature $T_{dark}$ is the temperature {\sl without} illumination (the value is undisclosed in the manuscript) rather than the same temperature at which the illuminated experiments were conducted. Fitting $R_{photo}$ to the Arrhenius equation gives $E_{a,photo} = 0.59$ eV ($0.67$ eV) for CW (white light) illumination, both smaller than the activation energy in the dark, $E_{a,dark} = 0.8$ eV. The authors conclude that the apparent reduction of activation energies is due to ``hot'' electrons. We note in passing that the extraction of exponents is based on a surprisingly limited number of data points and a narrow temperature range (see Appendix~\ref{app:data}), such that the differences between activation energies lie within the margin of error.

Already at this stage, it is evident that the thermal control experiment in~\cite{Halas-Nature-Catalysis-2020} is inadequate, since it does not account for the heating caused by the illumination. As a result, the activation energy extracted from $R_{photo}$ is misleading, since it includes the thermal contribution to the reaction rates. This practice is in direct contradiction of earlier work by some of the authors of the current manuscript~\cite{Halas_dissociation_H2_TiO2,Halas_Science_2018,seemala2019plasmon}, as well as their (and our) thermal calculations, and other recent work~\cite{Liu_thermal_vs_nonthermal,Liu-Everitt-Nano-Letters-2019}.

A more adequate thermal control experiment would be to measure the reaction rates while heating the sample to the same $T_S$ and account for the thermal gradients. However, as we demonstrate below, this control experiment conclusively shows that there is no contribution of non-thermal electrons to the reaction. 

To show this, we first obtain, using the data points of Ref.~\cite{Halas-Nature-Catalysis-2020} (black and green points), the total reaction rate under illumination, $R_{illum}$ (orange points in Fig.~\ref{fig:Corrected6b}). Next, to obtain the correct contribution from the photocatalysis ($R_{photo}$, red points in Fig.~\ref{fig:Corrected6b}), we subtract from $R_{illum}$ not the reaction rate in the dark (as was done in Ref.~\cite{Halas-Nature-Catalysis-2020}), nor the thermal reaction rate assuming a uniform temperature $T_S$ in the sample volume, but rather the thermal reaction rate due to a temperature distribution which has a surface temperature $T_S$ {\em identical} to the one measured as well as the associated thermal gradients (calculated in Appendix~\ref{app:sims}), namely, 
\begin{equation}\label{eq:R_correct}
R_{photo} \approx R_{illum}(T_S)- \int R_{dark}(T(\vec{r})) d\vec{r}.
\end{equation}
Here, $T(\vec{r})$ is the calculated temperature throughout the sample, and $R_{dark}(T) = R_0 \exp \left(-\frac{E_a}{k_B T(\vec{r})}\right)$ is the Arrhenius law, with $E_a\sim 0.8$eV extracted from the data in the dark (black points in Fig.~\ref{fig:Corrected6b}). 

The Arrhenius fit to Eq.~(\ref{eq:R_correct}) (red points in Fig.~\ref{fig:Corrected6b}) gives $E_{a,photo} \approx 0.7$ eV ($\approx 0.72$ eV) for illumination with CW (white light). Both changes are much smaller than the reported reduction; the obtained values are also well within the reported error bars, such that they cannot serve as conclusive evidence of any ``hot electron'' driven mechanism. Put simply, even if one assumes that there is ``hot-electron'' photocatalysis, its activation energy is roughly the same as the thermal one.

\subsection{A pure thermal analysis}
We have shown above that a re-examination of the control experiment and data analysis~\cite{Halas-Nature-Catalysis-2020} invalidates the conclusion of the paper. Now, we continue and show that a pure thermal analysis can explain the data perfectly.

This requires pointing out yet another possible flaw in~\cite{Halas-Nature-Catalysis-2020}, namely, the setting of the sample emissivity.
As in the authors' prior work~\cite{Halas_Science_2018}, a surprisingly high value ($0.95$~\cite[SI, p.~28]{Halas-Nature-Catalysis-2020}) was used, which we showed previously to be insufficiently accurate by the authors' own calibration experiment~\cite{Y2-eppur-si-riscalda,R2R}~\footnote{In particular, a difference of a few percent in the reading of the thermal camera and a thermocouple causes an uncertainty of hundreds of percent in the reaction rate.}. Moreover, not only this value is the {\em default} value of the camera, a calculation based on the composition of the sample in the current study (as done in~\cite{anti-Halas-Science-paper,R2R}) predicts an emissivity of $\approx 0.17$, significantly lower compared to the setting in the experiment~\footnote{It is also worth mentioning that our previous work~\cite{anti-Halas-Science-paper,R2R} showed conclusively that the IR camera was not operated properly in previous work of this group~\cite{Halas_Science_2018} (camera-sample distance was unrealistically high; image out of focus); see also discussion in~\cite{Baffou-Quidant-Baldi}. Since the authors do not provide any detail regarding the camera-to-sample distance, nor evidence for proper focusing, it is impossible to know if the camera was operated correctly in the current experiment. }.

Furthermore, a thermocouple embedded into the sample measured $10-20^\circ$C higher in the dark and up to $200^\circ$C lower under illumination compared to the thermal camera (see Figs.~S13(c) and S13(d), respectively; the former confirms the inaccuracy in the chosen emissivity). Remarkably, in~\cite[SI, p.~27]{Halas-Nature-Catalysis-2020}, the authors question the validity of the thermocouple reading, and disregard it. Their reasoning is that the disagreement between the thermocouple and camera readings results from the inability of the thermocouple to measure surface temperature ``due to the limited light penetration into the catalyst bed and the heat localization on the surface of the irradiated catalyst''. Not only does the second part of this statement contradict an underlying assumption of the analysis (of a uniform sample temperature), but in fact the entire argument is physically incorrect -- the limited light penetration does not prevent heat diffusion to the lower parts of the sample (see e.g.,~\cite{Un-Sivan-sensitivity}). The argument is also logically flawed -- the disagreement in the two temperature readings indicates that either one (or both) of the measurements is flawed, or that both are correct, but measure somewhat different quantities. 

It seems more than plausible that not only is the camera reading likely to be inaccurate, but also that the difference between the two temperature readings is to be expected. Indeed, there is no denying that vertical temperature gradients do exist in such heated or illuminated catalyst structures. Again, such gradients were measured by some of the authors of the current manuscript~\cite{seemala2019plasmon} and by others (e.g.~\cite{Meunier_T_uniformity,Liu-Everitt-Nano-Letters-2019}); they can also be extrapolated from the numerical simulations presented in Fig. S14 (performed over small voxels), and were described in detail in~\cite{Y2-eppur-si-riscalda,Un-Sivan-sensitivity} -- see also Appendix~\ref{app:sims}. 

Now, since the high emissivity setting may result in a temperature reading which is colder from the actual surface temperature, we now show that the data can be fully reproduced with only thermal reactions. To do so, we evaluate the reaction rate by integrating the Arrhenius law over the entire sample, taking into account the calculated gradients $T(\vec{r})$ (see Appendix~\ref{app:sims}) and assuming that the {\sl real} surface temperature is $T_S$ plus some temperature shift $\delta T$, i.e., $\int d\vec{r} R_0 \exp\left(-\frac{E_a}{k_B (T(\vec{r};T_S) + \delta T)}\right)$, with $R_0$ and $E_a$ taken from an Arrhenius fit to the reaction rate in the dark.

\begin{figure}[h]
\centering{
\includegraphics[width=0.75\textwidth]{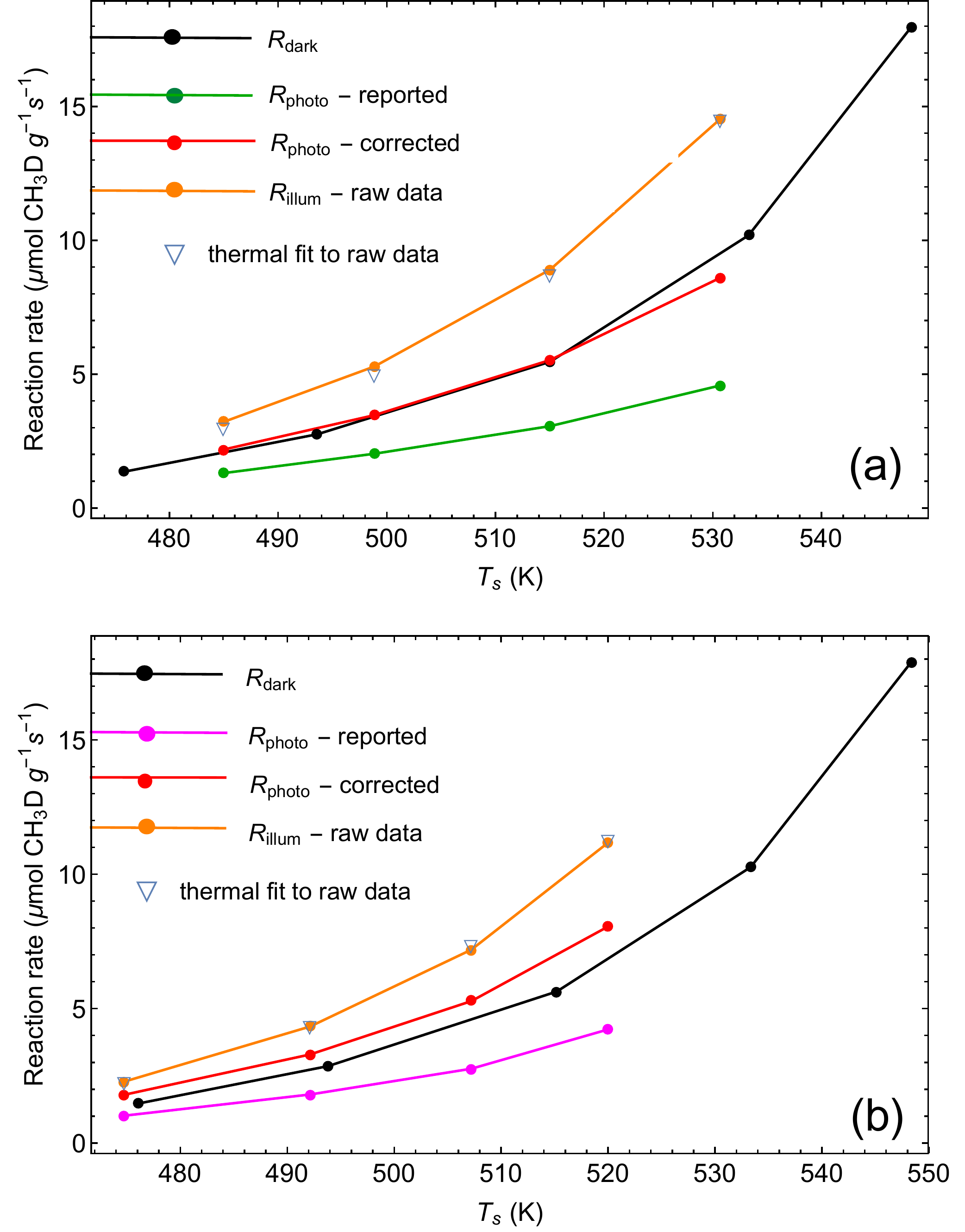}
\caption{Reaction rates for (a) CW and (b) white light illumination as a function of the measured surface temperature $T_S$. Black points: reaction rate in the dark, $R_{dark}(T_S)$, giving (from an Arrhenius fit) an activation energy of $0.8 eV$. green (a; CW) and magenta (b; white light) points: contribution of ``hot'' electrons to the reaction rate under illumination, $R_{photo} = R_{illum}(T_S) - R_{dark}(T_S)$, as evaluated in~\cite{Halas-Nature-Catalysis-2020}, disregarding the temperature difference between the bulk and surface reported by the authors themselves. Red points: corrected contribution of ``hot'' electrons to the reaction rate under illumination~(\ref{eq:R_correct}), with a much smaller change in the activation energy. Orange points: total reaction rate (black + green/pink points; raw data); Triangles: fit to raw data using a purely thermal model with a slightly shifted surface temperature; perfect agreement is observed.} \label{fig:Corrected6b}}
\end{figure}

In particular, if the temperatures are shifted by $\delta T = 27$K (CW) and $\delta T = 36$K (white light), one obtains {\em perfect} predictions of the measured reaction rate {\em without modifying the activation energy nor the prefactor} (empty triangles in Fig.~\ref{fig:Corrected6b}). Such a shift is of the same order of magnitude as the discrepancy between the thermocouple and camera readings reported by the authors themselves (Fig. S13(c)), and much smaller than the estimated error due to the emissivity and comparable to the independent thermal calculations shown in Appendix~\ref{app:sims}.

Thus, by Ockham's razor, it seems far more likely that our simple yet remarkably quantitative thermal argument is the correct explanation for the faster chemistry reported compared with the speculative explanation provided in the original manuscript. Definitive proof of ``hot'' electron driven catalysis, particularly in light of the issues discussed above, will require approaches that can more precisely isolate thermal effects.

\section{Absence of $^{13}$CH$_3$F in mass spectrometry results (Fig. 2b)}

Aside from the (problematic) analysis of the purported photocatalysis contribution to the reaction rate, the authors reported other results that they claim support involvement of ``hot'' electrons.  We find these observations and analyses unconvincing, especially as originally reported in~\cite{Halas-Nature-Catalysis-2020}, as described in the following.

The time evolution of the starting material CH$_3$F ($m/z$ 34) and desired product CH$_3$D ($m/z$ 17) as followed by mass spectrometry are shown in Fig.~2b of~\cite{Halas-Nature-Catalysis-2020}. To demonstrate the selectivity of the reaction, traces for several side products, including CH$_2$DF ($m/z$ 35), are also displayed and remain at zero. 
However, $^{13}$CH$_3$F should also appear with $m/z$ 35; in its $1.1\%$ natural abundance~\cite{nistch3fms}, this translates to $1.1\%$ of the reported $^{12}$CH$_3$F ($m/z$ 34) response or $(7.1)(0.011) = 0.078$, which should be visible on this plot, yet is inexplicably absent. Consider by contrast that care was taken to provide an explanation for the absence of the signal for \ce{DF}, but no such physical or chemical explanation is possible for the absence of $^{13}$CH$_3$F.

Even if we presume that the authors made an unannounced baseline correction, this type of presentation is unjustified here because these background levels are pertinent chemical data. If such a correction was performed here it was inconsistently applied, as some species (e.g.~DF in Fig.~3 of~\cite{Halas-Nature-Catalysis-2020}) are not baseline-corrected, and there is no basis to categorize species (to be more precise, $m/z$ values) as reactants vs.~products, since $m/z$ 35 is decidedly both. The $^{13}$CH$_3$F $m/z$ signal would decrease in roughly the same proportion as the $^{12}$CH$_3$F ($0.2/7.1 = 2.8\%$), which translates to changes of 220 counts/s in the $m/z$ 35 and 18 ($^{13}$CH$_3$D) channels. Based on the signal and noise levels shown in Fig.~S4 of~\cite{Halas-Nature-Catalysis-2020}, such changes would be near the observable limit if the data were plotted appropriately. It is unclear how the authors propose to distinguish between different species with the same $m/z$, which leads us to the larger problems with the mass spectrometry approach reported here.

Mass spectra require interpretation~\cite{fred1993interpretation}. Electron impact ionization causes severe fragmentation (see e.g., the NIST spectra of $^{12}$CH$_3$F cited above~\cite{nistch3fms}), and in this case it would be mostly coincidence if no daughter ions fell into the mass channels that the authors desired to interpret as products. Furthermore, this fragmentation varies substantially across isotopologues, both for statistical and other reasons -- discussion of this phenomenon in the relevant example of deuteromethanes can be found
in~\cite{dibeler1950mass}. Full mass range scans are essential to identifying the molecules present (if possible) and to check for the presence of other species (for example, methane or ethane in this case). Such information is especially important in the present work, where highly reactive species such as DF and methyl radicals are present, and only an extremely restricted reaction mechanism has been considered.

\section{Reaction order and stochiometry (Figs.~6b and 6c)}
Given the less than ideal execution and analysis of the mass spectrometry aspects of this study, its authors choose instead to highlight the overall stoichiometric carbon balance and the observed change in D$_2$ reaction order (Figs.~6b-c of~\cite{Halas-Nature-Catalysis-2020}).   However, the CH$_3$F and CH$_3$D curves do not mirror one another completely, indicating that the stoichiometry of the reaction evolves over the course of the illumination, such that the carbon balance and mechanism are not as straightforward as the authors would have us believe.

Regarding the reaction order change, while the change in reaction order for D$_2$ is noteworthy, such an effect could certainly be caused by a temperature increase, via the temperature dependence of the elementary step rate constants and equilibrium constants (See Supplementary Note 4 in the SM of \cite{Halas-Nature-Catalysis-2020}), but no control experiment was performed to test this possibility. Furthermore, the proposed explanation that illumination would enhance D$_2$ desorption is also what one would expect from elevated surface temperatures~\footnote{Parenthetically, the incident power density of 1.4 W/cm$^{-2}$ used in Fig.~6b is higher than the range of 0.6-1.1 W/cm$^{-2}$ given otherwise. }.

Finally, we would like to emphasize yet again that our purpose here is not to disprove contributions from photocatalytic mechanisms, but rather to argue that the experiments and analysis of Halas {\em et al.} as originally published in~\cite{Halas-Nature-Catalysis-2020} do not meet the burden of proof.

\appendix

\section{Data analysis in~\cite{Halas-Nature-Catalysis-2020}}\label{app:data}

In addition to the technical and conceptual problems described above, the original data analysis of the manuscript suffers from insufficient data recording.

Indeed, somewhat surprisingly, the original analysis relied on measurement of only 4 data points, does not incorporate any error bars on the experimental accuracy, and is performed over a limited temperature range of a few tens of degrees (compared to heating by many hundreds of degrees in~\cite{Halas_Science_2018}). As a result, the extracted $R_{photo}$ varies modestly (compared to changes by several orders of magnitude, see e.g.,~\cite{Halas_dissociation_H2_TiO2,plasmonic_photocatalysis_Linic,Halas_Science_2018}), making it highly sensitive to the details of the data analysis. This analysis consisted of, in addition to the obvious Arrhenius fit, an interpolation of the measured data. Indeed, since under illumination the temperature profile is different than in the thermocatalysis control experiment~\cite{Liu-Everitt-Nano-Letters-2019,Y2-eppur-si-riscalda,R2R} (in the current case, by several tens of degrees; data not shown), the correlation of the photocatalysis experiment to its control required proper lowering of the heater temperature. As a result, the data was eventually obtained at different surface temperatures for the photocatalysis and thermocatalysis control.

Thus, as discussed in~\cite{Y2-eppur-si-riscalda,R2R}, due to the exponential sensitivity inherent to the Arrhenius Law, the extraction of the activation energy from the data is associated with a very large error margin which makes any claims about the physics and chemistry underlying the reaction questionable.

\section{Thermal calculations}\label{app:sims}
\begin{figure}[tb]
\centering{
\includegraphics[width=1\textwidth]{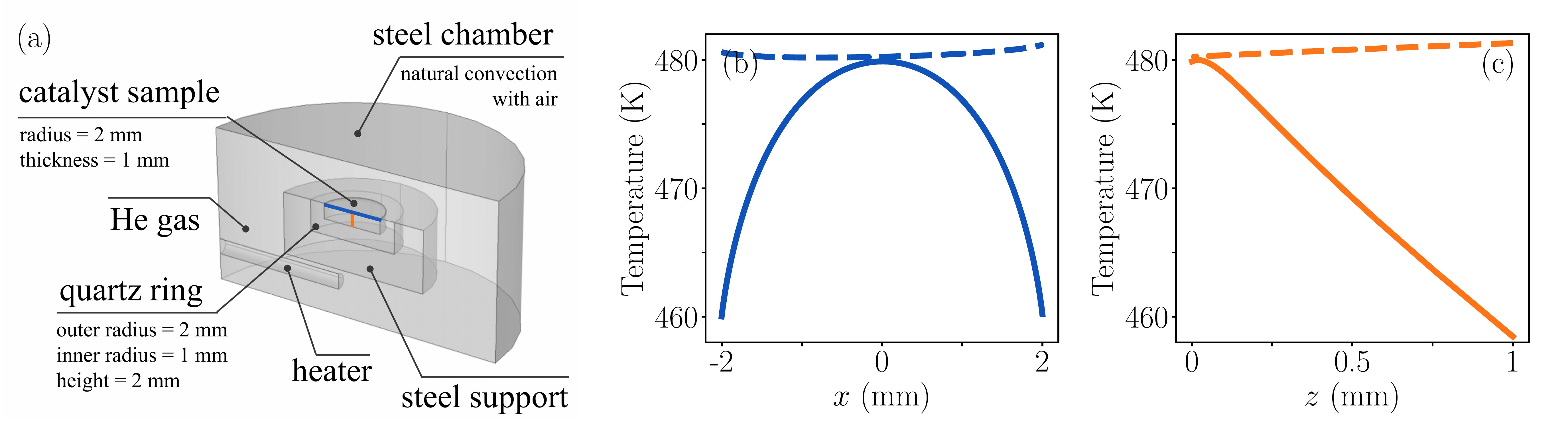}
\caption{(a) Details of the photocatalytic chamber used in~\cite{Halas-Nature-Catalysis-2020} which we used for reproducing the temperature distribution. The temperature along the (b) horizontal and (c)  vertical cross-sections (shown in (a)) for the photocatalysis experiment using the broadband visible light illumination (450-800 nm, 2.0 W/cm$^2$) (solid lines) and thermocatalysis control experiment (dashed lines). }
\label{fig:temp_profile}}
\end{figure}

The temperature profile in the experiment described in~\cite{Halas-Nature-Catalysis-2020} was determined by a detailed thermal calculation that accounted for the geometry described in the manuscript, together with the appropriate (temperature-dependent) material parameters, see Fig.~\ref{fig:temp_profile}. While it could be possible that some of these details differ from the actual experiment, extensive simulations show a modest sensitivity to all relevant parameters, in agreement with our previous extensive investigation of this very topic~\cite{Un-Sivan-sensitivity}. Gas flow, which was not accounted for in the simulations, was already shown to be a very weak effect~\cite[Appendix C]{Y2-eppur-si-riscalda}).

One can observe clear temperature gradients in both the transverse and vertical directions of a few tens of degrees. As a result, the Arrhenius Law predicts that regions of the catalyst at the bottom and edges of the catalyst sample contribute as much as $50\%$ of the top (hottest) point.

The experimental setup in~\cite{Halas-Nature-Catalysis-2020} has a rather limited capability to identify these gradients. First, the available transverse resolution is demonstrated in Fig. S15, where the pixels seem to be roughly $0.5$mm in size; this is in line with the camera manual which indicates an optimal transverse resolution of $690\mu$m~\cite{R2R}. Note, however, the incorrect claims on the transverse resolution made by the authors $\sim 100\mu$m resolution [SI, p. 6]. Thus, the authors are also unaware to their inability to identify these gradients experimentally.

Vertical gradients of up to $200^\circ$K were observed in an offline temperature calibration experiments (Fig. S13) performed on a thicker sample. Further thermal simulations show that such gradients may be obtained by focusing the incoming beam, a practice that was adopted in the measurement described in Fig. S13 (but mentioned only in private correspondence with the editor).


\end{document}